\documentclass[final,5p,times,twocolumn]{elsarticle}
\usepackage{lipsum,lineno,amssymb,amsmath,braket,bm,ulem,braket,hyperref}
\usepackage[dvipsnames]{xcolor}

\usepackage{stfloats}

\journal{Journal of High Energy Astrophysics}

\begin{document}

\begin{frontmatter}

\title{Impact of electron spectra on morphology of pulsar halos at ultra-high energies}
\author[1]{Ying-Ying Guo}
\author[1,2]{Qiang Yuan\corref{cor1}}
\ead{yuanq@pmo.ac.cn}

\address[1]{Key Laboratory of Dark Matter and Space Astronomy, Purple Mountain Observatory, Chinese Academy of Sciences, Nanjing 210023, China}    
\address[2]{School of Astronomy and Space Science, University of Science and Technology of China, Hefei 230026, China}

\cortext[cor1]{Corresponding author}

\begin{abstract}
The extended $\gamma$-ray halos around pulsars are unique probe of transportation of high-energy
electrons (and positrons) in vicinities of such pulsars. Observations of morphologies of several
such halos indicate that particles diffuse very slowly around pulsars, compared with that in the 
Milky Way halo. The energy-dependent morphologies are expected to be very important in studying 
the energy-dependence of the diffusion coefficient. In this work we point out that the spectrum of
high-energy electrons takes effect in shaping the $\gamma$-ray morphologies at the ultra-high-energy 
bands, and thus results in a degeneracy between the electron spectrum and the energy-dependence of 
the diffusion coefficient. The reasons for such a degeneracy include both the Klein-Nishina effect 
of the inverse Compton scattering and the curvature (if any) of the electron spectrum. 
It it thus necessary to take into account the spectral shape of electrons when deriving the
energy-dependence of diffusion coefficient using ultra-high-energy $\gamma$-ray measurements of 
extended pulsar halos.
\end{abstract}
\begin{keyword}
pulsar halos \sep diffusion propagation \sep energy losses 
\end{keyword}

\end{frontmatter}

\section{Introduction}
\label{intro}

Charged cosmic rays (CRs) are expected to propagate diffusively in the interstellar medium (ISM)
due to scattering with random magnetized turbulence \cite{1990acr..book.....B,2007ARNPS..57..285S}.
According to the resonant scattering theory between particles and the interstellar turbulence, the
rigidity-dependence of the diffusion coefficient follows a power-law, $D(R)\propto R^{\delta}$, 
given a power-law form of the spectral energy density of turbulence, $w(k)\propto k^{-2+\delta}$ 
where $k$ is the wave number of the magnetohydrodynamic waves. Fitting to the CR secondary-to-primary
ratios, the parameter $\delta$ is found to be about $0.3-0.6$ for energies below TeV/n
\cite{2017PhRvD..95h3007Y,2019SCPMA..6249511Y}. Further precise measurements of the 
secondary-to-primary ratios show that there are hardenings of such ratios at high energies
\cite{2021PhR...894....1A,2022SciBu..67.2162D}, suggesting possible changes of the slope of rigidity-dependence of the diffusion coefficient \cite{2023FrPhy..1844301M}.

The secondary-to-primary ratios reflect an average effect of the propagation of CRs in the Milky Way.
It is very likely that the propagation is spatially dependent. Such a scenario was proposed to
explain some observational results of CRs and diffuse $\gamma$ rays \cite{2012ApJ...752L..13T,
2012PhRvL.108u1102E,2016ApJ...819...54G,2018PhRvD..97f3008G}. Observations of extended $\gamma$-ray
halos surrounding middle-aged pulsars, which are expected to be produced by the inverse Compton
scattering (ICS) of high energy electrons (and positrons) produced by the pulsars with the 
interstellar radiation field, offer a more direct measurement of the particle diffusion in 
specific regions of the ISM \cite{2017Sci...358..911A,2021PhRvL.126x1103A}. The results show that 
electrons propagate much more slowly around pulsars than that in the Milky Way halo, indicating an
inhomogeneous propagation of particles \cite{2018ApJ...863...30F,2018PhRvD..97l3008P}. 
Additional supports for an inhomogeneous propagation were also given from the spectra of electrons 
and positrons, nuclei, and the large-scale anisotropies
\cite{2018PhRvD..98h3009H,2019JCAP...10..010L,2020ApJ...903...69F}. 

Detailed morphological studies of the $\gamma$-ray pulsar halos are very important in measuring
the energy-dependence of diffusion coefficient, and testing its generality in different places
of the Milky Way. Due to the fast energy losses of electrons in the ISM (synchrotron losses
and the ICS losses), higher energy electrons experience shorter lifetime which approximately
scales as $\tau(E_e)\propto E_e^{-1}$. Depending on the slope $\delta$ of the diffusion coefficient,
the extensions of $\gamma$-ray halos may increase (for $\delta>1$) or shrink (for $\delta<1$)
when energy increases. This expectation does not rely on the spectrum of electrons, and may hold 
for specific energy range where the cooling dominates the diffusion, the ICS lies well in the 
Thomson regime and the electron spectrum follows roughly a power-law.  However, in the
ultra-high-energy (UHE) range (with $\gamma$-ray energy $E_{\gamma}$ higher than tens of TeV)
many of the above assumptions are invalid, and this simple expectation should not hold any more
\cite{2022PhRvD.105j3007F}. In this work, we study the energy-dependent morphologies of pulsar 
halos in the UHE range assuming different spectral shapes. We will show that the electron spectrum 
couples with the diffusion process to shape the $\gamma$-ray morphologies and complicates the 
measurement of the energy-dependence of the diffusion coefficient. This effect cannot fe ignored 
since at UHE spectral cutoff is common for source acceleration.

\section{Injection and propagation of electrons from pulsars}

The propagation of electrons (and positrons) in the ISM around a pulsar is described by the
diffusion-cooling equation
\begin{equation}
\frac{\partial N}{\partial t} = \nabla \cdot (D \nabla N) + \frac{\partial}{\partial E_e} (b N) + Q,
\end{equation}
where $E_e$ is the energy of electrons, $N=N(r,E_e,t)$ is the differential density of electrons, 
$D=D(E_e)$ is the diffusion coefficient, $b=b(E_e)=-dE_e/dt$ is the energy loss rate, and 
$Q=Q(r,E_e,t)$ represents the source term. In this work we focuses on electron propagation near the 
pulsar, and neglect possible spatial variations of the diffusion coefficient and the energy loss rate.

The diffusion coefficient is parameterized as a power-law of electron energy, $D(E_e)=D_0E_e^{\delta}$.
The cooling of electrons includes mainly the synchrotron cooling and ICS cooling. The energy loss
rate of synchrotron radiation is 
\begin{equation}
b_{\rm syn}(E_e) = \frac{4}{3} \sigma_T c \gamma_e^2 U_B,
\end{equation}
where $\sigma_T$ is the Thomson cross section, $\gamma_e = E_e/(m_e c^2)$ is the Lorentz factor 
of electron, $m_e$ is the electron mass, $c$ is the speed of light, and $U_B = B^2 / (8 \pi)$ is 
the energy density of the magnetic field with strength $B$. In this work, $B=3~\mu$G is 
assumed\footnote{The magnetic field affects the relative weights between the synchrotron cooling
rate and the ICS cooling rate. A weaker magnetic field would result in more significant impacts
on the halo morphologies due to the Klein-Nishna effect and vise versa.}.
For the cooling rate due to the ICS emission, we use the approximation given in 
Ref.~\cite{2021ChPhL..38c9801F}. We assume three gray (black) body components of the interstellar 
radiation fields (ISRF), the cosmic microwave background with temperature of 2.725 K and energy 
density of 0.26 eV cm$^{-3}$, an infrared background with temperature of 20 K and energy density 
of 0.3 eV cm$^{-3}$, and an optical component with temperature of 5000 K and energy density of 
0.3 eV cm$^{-3}$, respectively. 

The injection function is assumed to be a point source in space, a power-law with (super)-exponential
cutoff in energy, and a profile following the pulsar spindown in time:
\begin{equation}
Q(\mathbf{r},E_e,t)=Q_0\delta(\mathbf{r}-\mathbf{r}_{\rm psr}) \times E_e^{-\alpha}\exp[-(E_e/E_c)^{\beta}]
\times (1+t/\tau_0)^{-2},
\end{equation}
where $Q_0$ is a normalization constant, $\alpha$ is the spectral index, $E_c$ is the characteristic 
cutoff energy of electrons, $\beta$ describes the sharpness of the cutoff, $\tau_0$ is the time 
scale that the spindown luminosity of the pulsar starts to decay. The parameter $\tau_0$ is difficult 
to be determined observationally, and we assume a nominal value of $10^4$ yr in this work.

The propagation equation (1) can be solved analytically using the Green's function method with 
respect to variables $\mathbf{r}$ and $t$ (i.e., $\delta$-function type injection for space and 
time), for an arbitrary injection spectrum \cite{1995PhRvD..52.3265A}. Convolving the Green's 
function with the injection time profile, one can obtain the electron distribution at give time 
and position. 

The photon emissivity due to the ICS is then given by
\begin{equation}
w_{\text{ICS}}(t, E_\gamma, \mathbf{r}) = \int_0^\infty d\epsilon \, n(\epsilon) \int_{E_{\text{min}}}^\infty 
dE_e \, N(t, E_e, \mathbf{r}) \, F(\epsilon, E_e, E_\gamma),
\end{equation}
where $n(\epsilon)$ is the number density of background photons with energy $\epsilon$, 
$N(t, E_e, \mathbf{r})$ is the electron density obtained via solving the propagation equation, 
and $F(\epsilon, E_e, E_\gamma)$ is the $\gamma$-ray yield function of the ICS according
to the Klein-Nishina cross section. Integrating $w_{\text{ICS}}$ along the line of sight, 
one can obtain the $\gamma$-ray surface brightness for specific energy $E_\gamma$.

\section{Results}
We take the Geminga pulsar halo as an example for the discussion. The distance of Geminga 
pulsar is adopted as $d_{\text{psr}}=250$ pc, and the age is $\tau_{\rm psr}=3.4\times10^5$ yr
\cite{2005AJ....129.1993M}. The diffusion coefficient is parameterized as a power-law of electron
energy, $D(E_e)=D_0(E_e/100\,{\rm TeV})^{\delta}$. As a benchmark, $D_0=10^{27}$ cm$^2$~s$^{-1}$
is assumed, which is several times smaller than that inferred by HAWC observations
\cite{2017Sci...358..911A}. A smaller value of the diffusion coefficient is to avoid the complexity 
of the superluminal propagation at very small scales \cite{2021PhRvD.104l3017R,2022ApJ...936..183B}, 
and does not loses generality of this work.

\begin{figure}[!htb]
\begin{center}
\includegraphics[width=0.5\textwidth]{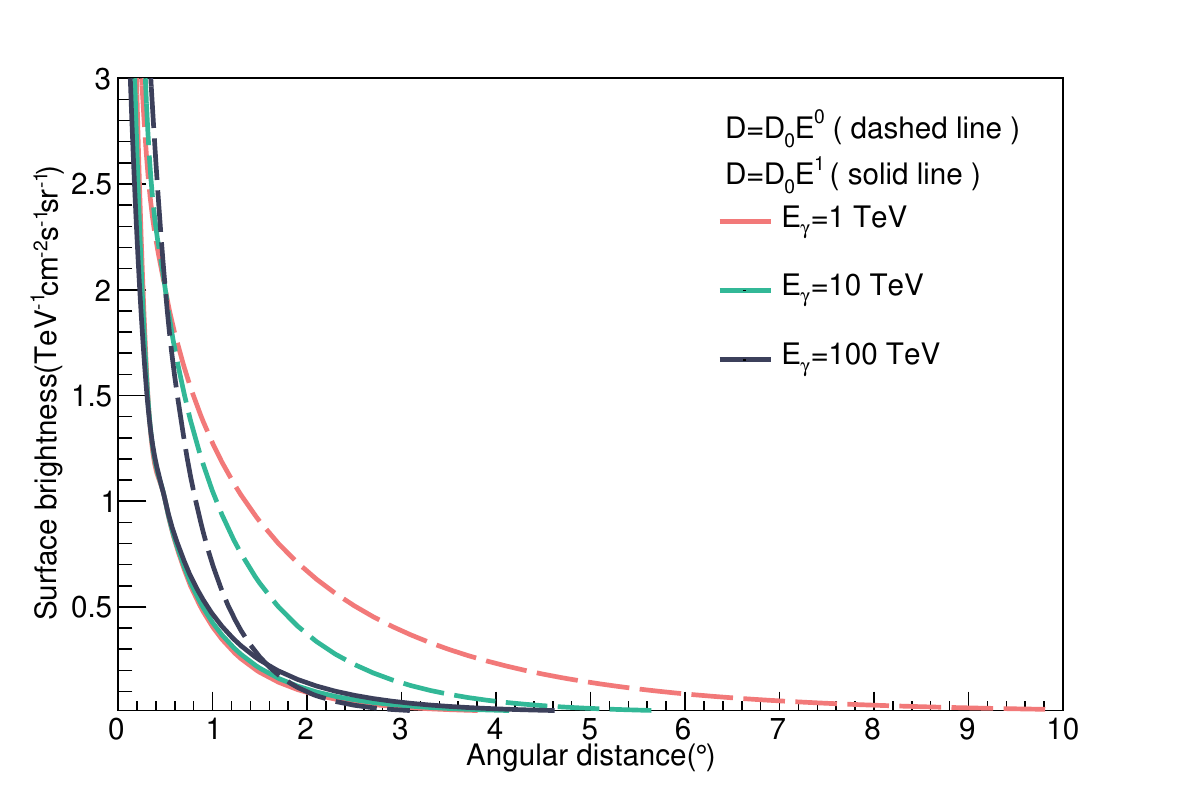}
\caption{Surface brightness distributions of $\gamma$ rays at different energies for $\delta=1$ 
(solid lines) and $\delta=0$ (dashed lines). The electron injection spectrum is a power-law 
form with spectral index $\alpha=2.5$.
}
\label{fig:delta_impact}
\end{center}
\end{figure}

Figure \ref{fig:delta_impact} illustrates the influence of $\delta$ on the evolution of the 
$\gamma$-ray morphology with energy. The electron injection spectrum is assumed to be a power-law 
function with index $\alpha=2.5$. Two values of $\delta$, 0 and 1 are studied. Dashed lines in this 
plot correspond to $\delta=0$, which show a clear contraction of the $\gamma$-ray extension with 
increasing energy. In this case, the energy loss rate of electrons increases with energy, but the 
diffusion coefficient keeps a constant, hence the characteristic diffusion distance 
$r_d\sim2\sqrt{D(E_e)\tau(E_e)}$ becomes smaller when energy is higher. For $\delta=1$, on the 
other hand, the characteristic diffusion length keeps almost unchanged, and the source extensions 
are similar at different energies as shown by solid lines. The extension at 100 TeV is slightly 
more extended than those at lower energies, due to that the ICS cooling rate becomes smaller and 
deviates from $E_e^{-1}$ thanks to the Klein-Nishina effect. 

\begin{figure*}[!htb]
\begin{center}
\includegraphics[width=1.0\textwidth]{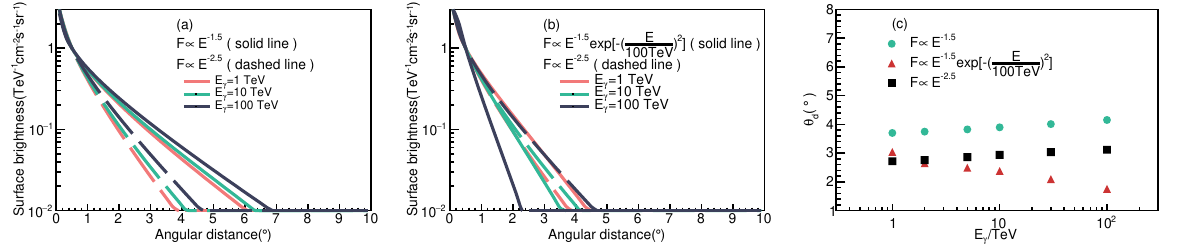}
\caption{Impact on the $\gamma$-ray morphologies from different electron injection spectra.
The energy-dependent slope of the diffusion coefficient is assumed to be $\delta=1$.
Panel (a): comparison of one-dimensional surface brightness distributions between power-law 
injection spectra with indices $\alpha=1.5$ (solid lines) and $\alpha=2.5$ (dashed lines). 
Panel (b): comparison of one-dimensional surface brightness distributions between power-law 
injection spectrum with index $\alpha=2.5$ (dashed lines) and power-law with super-exponential
cutoff spectrum $E_e^{-1.5}\exp[-(E_e/100\,{\rm TeV})^2]$.
Panel (c): variation of the diffusion angle $\theta_d$ with $\gamma$-ray energy for different 
injection spectra.
}
\label{fig:inject_impact}
\end{center}
\end{figure*}

Figure \ref{fig:inject_impact} shows the impact on the $\gamma$-ray morphologies from different
injection spectra of electrons, for $\delta=1$. Panel (a) shows the one-dimensional surface
brightness distributions for power-law injection spectra with indices $\alpha=1.5$ (solid lines) 
and $\alpha=2.5$ (dashed lines). Panel (b) of Figure \ref{fig:inject_impact} illustrates the 
difference between the power-law injection spectrum (with $\alpha=2.5$; dashed lines) and the 
power-law with super-exponential cutoff spectrum ($\alpha=1.5$, $E_c=100$ TeV, $\beta=2$; solid lines). 
To quantify the evolution of source extension, we fit the one-dimensional surface brightness
distribution with the following profile \cite{2017Sci...358..911A}
\begin{equation}
f(\theta)\propto\frac{1}{\theta_d(\theta+0.06\theta_d)}\cdot\exp\left[-(\theta/\theta_d)^2\right],
\end{equation}
where $\theta_d=r_d/d_{\rm psr}$ is the characteristic diffusion angle, and $\theta$ is the 
angular distance from the central pulsar. 
We assume $10\%$ relative errors of the surface brightness profiles for each angular bin 
and adopt a minimum $\chi^2$ method to do the fit. The evolution of $\theta_d$ for the three 
injection spectra are shown in panel (c). We can see that $\theta_d$ shows an increase tend when 
the energy increases for power-law injection, but shows a clear decrease trend for the spectrum 
with cutoff. It can also be noted that for $\alpha=1.5$, the $\gamma$-ray profiles are typically 
more extended than the case of $\alpha=2.5$. 
All these results show that the high-energy electron spectrum plays an important role in shaping 
the $\gamma$-ray morpholigies.

\begin{figure}[!htb]
\begin{center}
\includegraphics[width=0.5\textwidth]{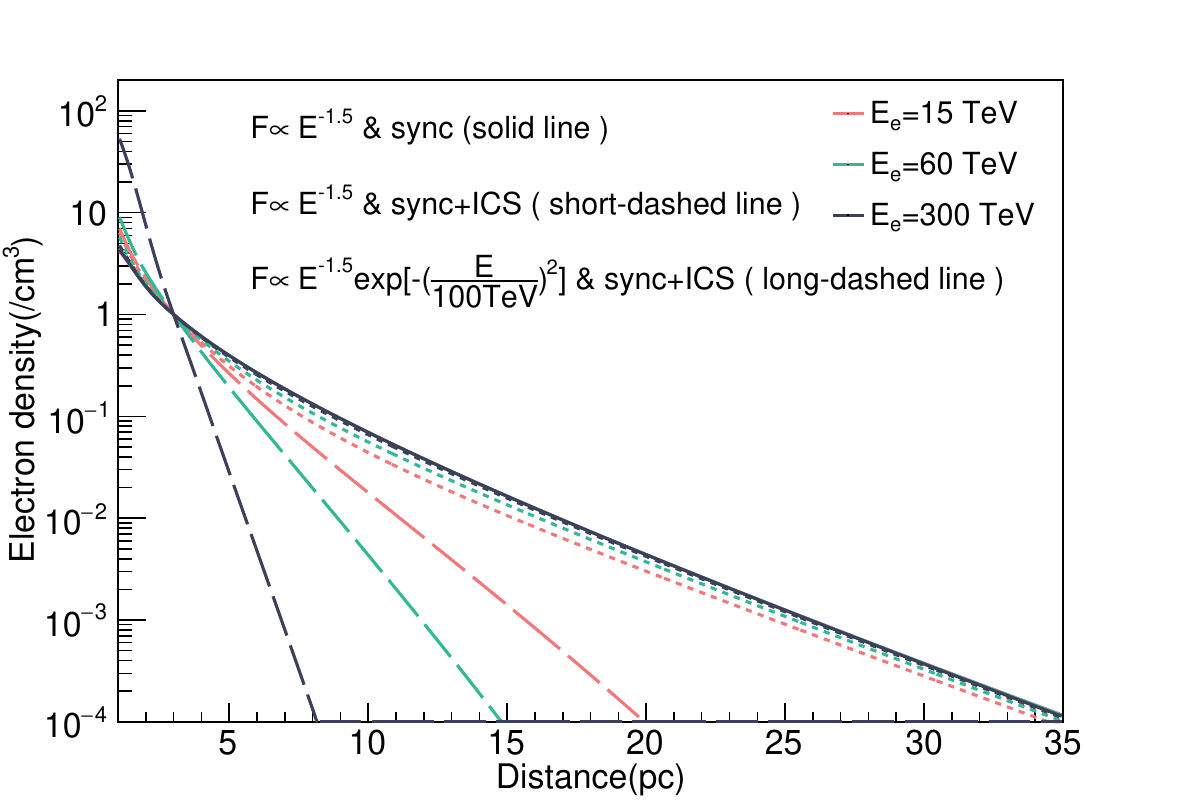}
\caption{Impact on the electron morphologies with respect energy loss rate and injection spectra.
The energy-dependent slope of the diffusion coefficient is assumed to be $\delta=1$. All curves
are normalized at 3 pc. In case of only synchrotron cooling, the radial profiles of electrons of different energies overlap with each other.}
\label{fig:electron_spatial}
\end{center}
\end{figure}

To better understand these results, we scrutinize the distributions of electrons at different
energies in detail. Figure \ref{fig:electron_spatial} show the radial profiles of electron 
density as a function of distance from the pulsar, for different post-propagation energies
and injection spectra. To eliminate the Klein-Nishina effect of the ICS, we test the case with
synchrotron cooling only for a power-law injection, as shown by the solid lines. The results
show that the spatial distributions of electrons with different energies are identical, just 
as expected for a cooling rate of $E_e^{-1}$ and $\delta=1$. However, when the ICS cooling is 
included (short-dashed lines), the cooling of higher energy electrons is smaller than $E_e^{-1}$, 
leading to more extended distributions of electrons with higher energies. As a result, the 
$\gamma$-ray morphology would also become more extended at higher energies, as shown in Figure
\ref{fig:inject_impact}. This also explains the morphology for harder spectrum ($\alpha=1.5$) 
is more extended than that for softer spectrum ($\alpha=2.5$), since there are more high-energy 
electrons for a harder spectrum. When there is cutoff of the injection spectrum, we in turn
observe that higher energy electrons are less extended than low energy ones (long-dashed lines).
In this case there are not many electrons at the source with energy high enough to cool down
to given energy (e.g., $E_e=300$ TeV). 
This could also be understood as that, under the continuous injection scenario, 
the current electron density is the time integral of the previously injected electron with 
higher energies. The decrease of high-energy electron flux for the cutoff spectrum is equivalent 
to a shorter effective integration time. The increase of the diffusion coefficient with energy 
($\sim E_e^1$) does not compensate the decrease in time, resulting in a decrease in $\theta_d$.
Therefore, we would observe a shrink of the $\gamma$-ray morphology with increasing energy if 
the source spectrum has a cutoff.

\begin{figure}[!htb]
\begin{center}
\includegraphics[width=0.5\textwidth]{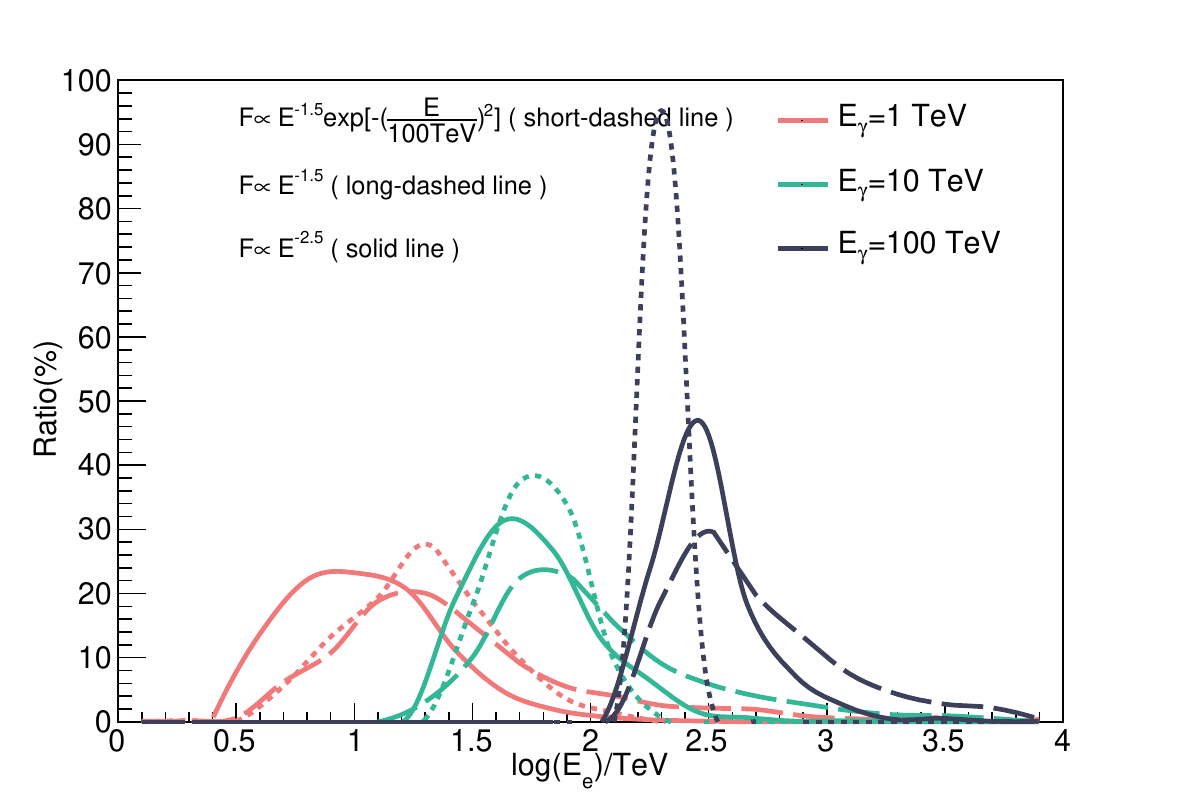}
\caption{The fraction distribution of injection electron energies that produce $\gamma$ rays 
with fixed energy $E_{\gamma}$.}
\label{fig:electron_energy}
\end{center}
\end{figure}

Finally, we show in Figure \ref{fig:electron_energy} the energy distribution of electrons which 
produce $\gamma$-rays with fixed energy $E_{\gamma}$, for different injection spectra.
Due to the wide energy distributions of the ISRF and the ICS cross section, the electron energy
distribution to give fixed $E_{\gamma}$ is quite wide. For power-law injections, a harder spectrum
corresponds to a longer high-energy tail of the electron energy distribution, which explains the
larger extension of the $\gamma$-ray morphology as we discussed in above. 
The $\gamma$-ray morphologies from 1 TeV to 10 TeV evolve slowly with energy (see panel (c) of 
Figure \ref{fig:inject_impact}), because the energies of electrons producing those $\gamma$ rays
are mainly below 100 TeV, and the Klein-Nishina effect of the cooling is not evident. However, 
from 10 TeV to 100 TeV, the Klein-Nishina effect becomes important, and the $\gamma$-ray 
extensions become significantly larger. 
For the injection spectrum with a curved shape, we can see that electron energy distributions
are more concentrated than the power-law injection cases. Even if we observe the same morphology
evolution of $\gamma$ rays, the inferred energy-dependent diffusion properties are different
for different injection spectra.

\section{Conclusion and Discussion}

Pulsar halos are formed by high-energy electrons and positrons injected from the associated
pulsar wind nebulae via the ICS off the background photons 
\cite{2004vhec.book.....A,2020A&A...636A.113G,2022FrASS...922100F,2022IJMPA..3730011L}. 
The $\gamma$-ray morphologies of pular halos are very important in revealing the particle 
propagation in the ISM around pulsars. It has been expected that the extension of the halo 
is proportional to the characteristic diffusion length scale, 
$\theta_d \propto r_d \propto 2\sqrt{D(E_e)\tau(E_e)}$. However, this expectation is 
over-simplified. In reality, the Klein-Nishina effect of the ICS cooling and the deviation 
of source injection spectrum from a power-law distribution affect the electron propagation 
and final distribution. As a consequence, the $\gamma$-ray morphology evolution also depends 
on the injection electron spectrum. 

This study investigates the impacts of the electron injection spectrum on the $\gamma$-ray 
morphology evolution of pulsar halos. For simplicity, we take the energy-dependent slope of 
the diffusion coefficient $\delta=1$ as a benchmark for discussion. Given the $E_e^{-1}$ 
energy loss rate of electrons due to synchrotron and ICS cooling, this diffusion coefficient 
predicts energy-independent $\gamma$-ray morphologies. It has been shown that at relatively 
low energies (e.g., $E_{\gamma}$ is smaller than tens of TeV), power-law injection spectrum 
indeed gives energy-independent source morphologies as expected. However, if $E_{\gamma}$ is 
as high as tens of TeV, even power-law injection spectrum results in an evolution of the 
source extension. This is because the Klein-Nishina effect of the ICS leads to a less efficient 
cooling effect than the $E_e^{-1}$ form. For power-law with super-exponential cutoff spectrum, 
a clear source contraction with increasing energy is given. We point out that such a contraction 
is due to that there are not enough high-energy electrons can contribute to those high-energy
photons because of the spectral cutoff. It is thus necessary to take into account the injection 
spectrum shape when studying the energy-dependence of the diffusion coefficient using the UHE 
$\gamma$-ray observations.

\section*{Acknowledgments}
This work is supported by the National Natural Science Foundation of China (No. 12220101003),
the Project for Young Scientists in Basic Research of Chinese Academy of Sciences (No. YSBR-061), 
and the Jiangsu Provincial Excellent Postdoctoral Program with grant number 2022ZB472.


\bibliographystyle{unsrt}
\bibliography{refs}

\begin{thebibliography}{10}

\bibitem{1990acr..book.....B}
V.~S. {Berezinskii}, S.~V. {Bulanov}, V.~A. {Dogiel}, and V.~S. {Ptuskin}.
\newblock {\em {Astrophysics of cosmic rays}}.
\newblock Amsterdam: North-Holland, 1990, edited by Ginzburg, V.L., 1990.

\bibitem{2007ARNPS..57..285S}
Andrew~W. {Strong}, Igor~V. {Moskalenko}, and Vladimir~S. {Ptuskin}.
\newblock {Cosmic-Ray Propagation and Interactions in the Galaxy}.
\newblock {\em Annual Review of Nuclear and Particle Science}, 57(1):285--327,
  November 2007.

\bibitem{2017PhRvD..95h3007Y}
Q.~{Yuan}, S.-J. {Lin}, K.~{Fang}, and X.-J. {Bi}.
\newblock {Propagation of cosmic rays in the AMS-02 era}.
\newblock {\em \prd}, 95(08):083007, April 2017.

\bibitem{2019SCPMA..6249511Y}
Q.~{Yuan}.
\newblock {Implications on cosmic ray injection and propagation parameters from
  Voyager/ACE/AMS-02 nucleus data}.
\newblock {\em Science China Physics, Mechanics, and Astronomy}, 62:49511,
  April 2019.

\bibitem{2021PhR...894....1A}
M.~{Aguilar}, L.~{Ali Cavasonza}, G.~{Ambrosi}, L.~{Arruda}, N.~{Attig},
  F.~{Barao}, L.~{Barrin}, A.~{Bartoloni}, S.~{Ba{\c{s}}e{\u{g}}mez-du Pree},
  J.~{Bates}, and et~al.
\newblock {The Alpha Magnetic Spectrometer (AMS) on the international space
  station: Part II - Results from the first seven years}.
\newblock {\em \physrep}, 894:1--116, February 2021.

\bibitem{2022SciBu..67.2162D}
F.~{Alemanno}, Q.~{An}, P.~{Azzarello}, F.~{Carla Tiziana Barbato},
  P.~{Bernardini}, X.~J. {Bi}, M.~S. {Cai}, E.~{Casilli}, E.~{Catanzani},
  J.~{Chang}, et~al.
\newblock {Detection of spectral hardenings in cosmic-ray boron-to-carbon and
  boron-to-oxygen flux ratios with DAMPE}.
\newblock {\em Science Bulletin}, 67(21):2162--2166, November 2022.

\bibitem{2023FrPhy..1844301M}
Peng-Xiong {Ma}, Zhi-Hui {Xu}, Qiang {Yuan}, Xiao-Jun {Bi}, Yi-Zhong {Fan},
  Igor~V. {Moskalenko}, and Chuan {Yue}.
\newblock {Interpretations of the cosmic ray secondary-to-primary ratios
  measured by DAMPE}.
\newblock {\em Frontiers of Physics}, 18(4):44301, August 2023.

\bibitem{2012ApJ...752L..13T}
N.~{Tomassetti}.
\newblock {Origin of the Cosmic-Ray Spectral Hardening}.
\newblock {\em \apjl}, 752:L13, June 2012.

\bibitem{2012PhRvL.108u1102E}
C.~{Evoli}, D.~{Gaggero}, D.~{Grasso}, and L.~{Maccione}.
\newblock {Common Solution to the Cosmic Ray Anisotropy and Gradient Problems}.
\newblock {\em \prl}, 108(21):211102, May 2012.

\bibitem{2016ApJ...819...54G}
Y.-Q. {Guo}, Z.~{Tian}, and C.~{Jin}.
\newblock {Spatial-dependent Propagation of Cosmic Rays Results in the Spectrum
  of Proton, Ratios of P/P, and B/C, and Anisotropy of Nuclei}.
\newblock {\em \apj}, 819:54, March 2016.

\bibitem{2018PhRvD..97f3008G}
Y.-Q. {Guo} and Q.~{Yuan}.
\newblock {Understanding the spectral hardenings and radial distribution of
  Galactic cosmic rays and Fermi diffuse {$\gamma$} rays with
  spatially-dependent propagation}.
\newblock {\em \prd}, 97(6):063008, March 2018.

\bibitem{2017Sci...358..911A}
A.~U. {Abeysekara}, A.~{Albert}, R.~{Alfaro}, C.~{Alvarez}, J.~D.
  {{\'A}lvarez}, R.~{Arceo}, J.~C. {Arteaga-Vel{\'a}zquez}, D.~{Avila Rojas},
  H.~A. {Ayala Solares}, A.~S. {Barber}, and et~al.
\newblock {Extended gamma-ray sources around pulsars constrain the origin of
  the positron flux at Earth}.
\newblock {\em Science}, 358:911--914, November 2017.

\bibitem{2021PhRvL.126x1103A}
F.~{Aharonian}, Q.~{An}, {Axikegu}, L.~X. {Bai}, Y.~X. {Bai}, Y.~W. {Bao},
  D.~{Bastieri}, X.~J. {Bi}, Y.~J. {Bi}, H.~{Cai}, and et~al.
\newblock {Extended Very-High-Energy Gamma-Ray Emission Surrounding PSR J 0622
  +3749 Observed by LHAASO-KM2A}.
\newblock {\em \prl}, 126(24):241103, June 2021.

\bibitem{2018ApJ...863...30F}
K.~{Fang}, X.-J. {Bi}, P.-F. {Yin}, and Q.~{Yuan}.
\newblock {Two-zone Diffusion of Electrons and Positrons from Geminga Explains
  the Positron Anomaly}.
\newblock {\em \apj}, 863:30, August 2018.

\bibitem{2018PhRvD..97l3008P}
Stefano {Profumo}, Javier {Reynoso-Cordova}, Nicholas {Kaaz}, and Maya
  {Silverman}.
\newblock {Lessons from HAWC pulsar wind nebulae observations: The diffusion
  constant is not a constant; pulsars remain the likeliest sources of the
  anomalous positron fraction; cosmic rays are trapped for long periods of time
  in pockets of inefficient diffusion}.
\newblock {\em \prd}, 97(12):123008, June 2018.

\bibitem{2018PhRvD..98h3009H}
Dan {Hooper} and Tim {Linden}.
\newblock {Measuring the local diffusion coefficient with H.E.S.S. observations
  of very high-energy electrons}.
\newblock {\em \prd}, 98(8):083009, October 2018.

\bibitem{2019JCAP...10..010L}
Wei {Liu}, Yi-Qing {Guo}, and Qiang {Yuan}.
\newblock {Indication of nearby source signatures of cosmic rays from energy
  spectra and anisotropies}.
\newblock {\em \jcap}, 10(10):010, Oct 2019.

\bibitem{2020ApJ...903...69F}
Kun {Fang}, Xiao-Jun {Bi}, and Peng-Fei {Yin}.
\newblock {DAMPE Proton Spectrum Indicates a Slow-diffusion Zone in the nearby
  ISM}.
\newblock {\em \apj}, 903(1):69, November 2020.

\bibitem{2022PhRvD.105j3007F}
Kun {Fang} and Xiao-Jun {Bi}.
\newblock {Interpretation of the puzzling gamma-ray spectrum of the Geminga
  halo}.
\newblock {\em \prd}, 105(10):103007, May 2022.

\bibitem{2021ChPhL..38c9801F}
Kun {Fang}, Xiao-Jun {Bi}, Su-Jie {Lin}, and Qiang {Yuan}.
\newblock {Klein-Nishina Effect and the Cosmic Ray Electron Spectrum}.
\newblock {\em Chinese Physics Letters}, 38(3):039801, March 2021.

\bibitem{1995PhRvD..52.3265A}
A.~M. {Atoyan}, F.~A. {Aharonian}, and H.~J. {V{\"o}lk}.
\newblock {Electrons and positrons in the galactic cosmic rays}.
\newblock {\em \prd}, 52:3265--3275, September 1995.

\bibitem{2005AJ....129.1993M}
R.~N. {Manchester}, G.~B. {Hobbs}, A.~{Teoh}, and M.~{Hobbs}.
\newblock {The Australia Telescope National Facility Pulsar Catalogue}.
\newblock {\em \aj}, 129:1993--2006, April 2005.

\bibitem{2021PhRvD.104l3017R}
S.~{Recchia}, M.~{Di Mauro}, F.~A. {Aharonian}, L.~{Orusa}, F.~{Donato},
  S.~{Gabici}, and S.~{Manconi}.
\newblock {Do the Geminga, Monogem and PSR J0622+3749 {\ensuremath{\gamma}}
  -ray halos imply slow diffusion around pulsars?}
\newblock {\em \prd}, 104(12):123017, December 2021.

\bibitem{2022ApJ...936..183B}
Li-Zhuo {Bao}, Kun {Fang}, Xiao-Jun {Bi}, and Sheng-Hao {Wang}.
\newblock {Slow Diffusion is Necessary to Explain the {\ensuremath{\gamma}}-Ray
  Pulsar Halos}.
\newblock {\em \apj}, 936(2):183, September 2022.

\bibitem{2004vhec.book.....A}
Felix~A. {Aharonian}.
\newblock {\em {Very high energy cosmic gamma radiation : a crucial window on
  the extreme Universe}}.
\newblock World Scientific Publishing, 2004.

\bibitem{2020A&A...636A.113G}
G.~{Giacinti}, A.~M.~W. {Mitchell}, R.~{L{\'o}pez-Coto}, V.~{Joshi}, R.~D.
  {Parsons}, and J.~A. {Hinton}.
\newblock {Halo fraction in TeV-bright pulsar wind nebulae}.
\newblock {\em \aap}, 636:A113, April 2020.

\bibitem{2022FrASS...922100F}
Kun {Fang}.
\newblock {Gamma-ray pulsar halos in the Galaxy}.
\newblock {\em Frontiers in Astronomy and Space Sciences}, 9:1022100, October
  2022.

\bibitem{2022IJMPA..3730011L}
Ruo-Yu {Liu}.
\newblock {The physics of pulsar halos: Research progress and prospect}.
\newblock {\em International Journal of Modern Physics A}, 37(22):2230011,
  August 2022.

\end{thebibliography}

\end{document}